\documentclass[aps,prb,reprint,amsmath]{revtex4-2}
\usepackage{setspace} 
\usepackage{hyperref}
\usepackage{color}
\usepackage[caption=false]{subfig}
\usepackage{graphicx,epstopdf}
\usepackage{bm}
\usepackage{lipsum}

\usepackage{float}

\begin{document}
\title{{\Large{}{}Universal Dynamics of a Spin-$3/2$ Fermi Gas in Traps with Distinct Spectra}}

\author{Shuyi Li}

\affiliation{Institute of Theoretical Physics, University of Science and Technology Beijing, Beijing 100083, China}
\author{Qiang Gu}
\email[Corresponding author: ] {qgu@ustb.edu.cn}
\affiliation{Institute of Theoretical Physics, University of Science and Technology Beijing, Beijing 100083, China}

\begin{abstract}

Universal power-law decay of spin-mixing oscillations has been observed in a harmonically trapped spin-$3/2$ Fermi gas, whose exactly equally spaced spectrum represents a highly special case. To gain insight into the origin of this behavior, we investigate the dynamics in traps with increasing and decreasing level spacings, represented by the infinite square well and the Pöschl--Teller potential, respectively. Despite pronounced differences in their single-particle spectra, all systems exhibit power-law decay of coherent oscillations, described by $A(t) = A_{0} - \gamma t^{\alpha}$. The exponent $\alpha$ remains insensitive to particle number, interaction strength, and the overall energy scale, but exhibits systematic differences among traps with distinct spectral structures. In contrast, the decay parameter $\gamma$ is strongly influenced by both the spectral structure and the overall energy scale. These findings demonstrate that power-law decay is not restricted to the harmonic trap but persists across qualitatively distinct spectral structures. The observed variation of $\alpha$ among different traps further suggests that the underlying single-particle spectrum plays an important role in shaping the decay dynamics.
 

\end{abstract}

\maketitle

\section{Introduction}

Universal behavior is a central concept in physics, describing the emergence of scaling laws that are largely independent of microscopic details. Originally established in equilibrium critical phenomena within the framework of renormalization-group theory\cite{K.G.Wilson-1975,K.G.Wilson-1983}, the concept of universality has recently been extended to far-from-equilibrium quantum dynamics, where robust scaling behavior has been observed in a variety of interacting many-body systems\cite{Erne-2018,Prufer-2018,Pecci-2022,Rodriguez-2022,ZhaoyiLi-2023,LiY-2024}. Ultracold atomic gases provide an ideal platform for exploring such phenomena because of their high controllability and tunability\cite{Chin-2010,Bloch-2008}, as well as their suitability for studying nonequilibrium dynamics\cite{Pecci-2022,Polkovnikov-2011,Ueda-2020}. In trapped systems, the single-particle spectrum is directly determined by the geometry of the confining potential\cite{Stringari-1996,Vignolo-2000,Minguzzi-2002}. Since the energy spectrum is closely related to phase evolution and quantum revival phenomena in isolated systems\cite{Berry-1977,Robinett-2004}, confinement geometry provides a natural route for investigating how spectral properties influence nonequilibrium dynamics.

In particular, interacting high-spin Fermi gases exhibit rich spin-mixing dynamics enabled by their additional spin degrees of freedom and intrinsic spin-exchange interactions\cite{Krauser-2014,Dong-2013,FengT-2017,Lisy-2025,Lisy-2026}. Unlike conventional spin-$1/2$ Fermi gases, high-spin systems support intrinsic coherent spin-exchange processes between different Zeeman components. Such nonequilibrium spin dynamics has been experimentally observed in ultracold large-spin $^{40}$K Fermi gases, stimulating extensive studies of collective spin oscillations, spin waves, and relaxation phenomena in multicomponent fermionic systems\cite{Krauser-2012,Heinze-2013,Ebling-2014}. Among these systems, spin-$3/2$ fermions constitute the minimal platform supporting intrinsic spin mixing while retaining a relatively simple theoretical structure, making them particularly suitable for exploring nonequilibrium many-body dynamics.

Recent studies have shown that spin-mixing oscillations in harmonically trapped spin-$3/2$ Fermi gases exhibit a universal power-law decay\cite{Lisy-2025}. However, that work focused exclusively on the harmonic oscillator trap (HOT), whose exactly equally spaced single-particle spectrum represents a highly special case. Consequently, despite the observation of such power-law decay, its physical origin remains poorly understood. In particular, it is unclear whether the observed power-law decay is a specific consequence of harmonic confinement or a more general feature of nonequilibrium spin dynamics. Motivated by this question, we extend the study to trapping potentials with qualitatively different spectral properties. Specifically, we consider the infinite square well (ISW), characterized by increasingly separated level spacings, and the Pöschl--Teller potential (PTP), which exhibits progressively compressed level spacings\cite{Poschl-1933,Flugge-1994,Lekner-2007,daSilva-2024,Ahmed-2001}. By comparing these distinct trapping potentials, we investigate the robustness of the power-law decay and examine how its decay characteristics depend on the underlying single-particle spectral structure.

To enable a meaningful comparison, all models are calibrated so that the lowest excitation energy spacing is identical, thereby isolating the influence of spectral properties on the many-body dynamics. Furthermore, by varying the trap width, we separately investigate the influence of the overall energy scale, corresponding to a global rescaling of the single-particle spectrum. Our results show that the decay of coherent oscillations universally follows a power-law form across all trapping potentials considered. The decay exponent $\alpha$ remains insensitive to particle number, interaction strength, and the overall energy scale, but varies systematically among the three traps. Specifically, the PTP exhibits the largest exponent ($\alpha\approx0.878$), the HOT an intermediate value ($\alpha\approx0.767$), and the ISW the smallest exponent ($\alpha\approx0.653$). In contrast, the decay parameter $\gamma$ depends strongly on both the spectral properties and the overall energy scale. These results demonstrate that the observed power-law decay is not an accidental consequence of the harmonic confinement, but rather a robust feature of nonequilibrium spin dynamics across qualitatively distinct spectral structures. Moreover, the systematic variation of the decay exponent among different traps suggests that the underlying single-particle spectrum plays an important role in shaping the damping dynamics, providing a possible route for controlling the decay dynamics in trapped high-spin Fermi gases.

\section{Model and method}

\begin{figure*}
	\begin{center}
		\includegraphics[width=0.9\linewidth]{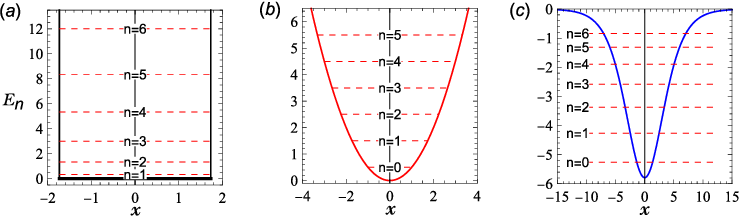}
	\end{center}
	\caption{\label{potential} {\bf Schematic illustration of three one-dimensional traps and their corresponding single-particle energy spectra.} (a) infinite square well (ISW) with $L_{\mathrm{ISW}} = 3.84$, (b) harmonic oscillator trap (HOT), and (c) P\"oschl--Teller potential (PTP). The horizontal dashed lines indicate the lowest few energy levels $En$. The first level spacing is fixed to $\Delta E_1 = 1$ in all three cases. The parameters used for visualization are $\lambda = 10$ and $a_{\mathrm{PTP}} = 0.324$ (corresponding to $V_0 = 5.789$) for the PTP. These parameters are chosen for illustrative purposes to clearly display only a few low-lying levels. In the actual calculations, a larger value $\lambda = 60$ is adopted, so that the number of bound states exceeds the number of occupied levels and finite-size effects are negligible.}
\end{figure*}

To investigate the role of spectral structure in nonequilibrium dynamics, we consider three paradigmatic one-dimensional traps: the infinite square well (ISW), the harmonic oscillator trap (HOT), and the P\"oschl--Teller potential (PTP). As illustrated in figure \ref{potential}, these systems possess qualitatively distinct single-particle spectra. For clarity, only several low-lying energy levels are shown schematically. The ISW [figure \ref{potential}(a)] is characterized by increasing level spacings, the HOT [figure \ref{potential}(b)] exhibits an exactly equally spaced spectrum, and the PTP [figure \ref{potential}(c)] features progressively compressed level spacings. This classification provides a convenient framework for analyzing how different spectral structures influence many-body dynamics.

The dimensionless Hamiltonian of the one-dimensional infinite square well (ISW) is given by  
\begin{equation}
	H = -\frac{1}{2}\frac{\mathrm{d}^{2}}{\mathrm{d}x^{2}},
	\qquad |x|<\frac{L}{2},
	\label{eq:H_ISW}
\end{equation}
where $L$ represents the width of the well. The normalized wavefunctions are
\begin{equation}
	\psi_n(x)
	= \sqrt{\frac{2}{L}}\,
	\sin\!\left[\frac{n\pi}{L}\left(x+\frac{L}{2}\right)\right].
	\label{eq:psi}
\end{equation}
The corresponding energy spectrum is $E_n = \frac{1}{2}\left(\frac{n\pi}{L}\right)^2$, indicating that the level spacing increases linearly with the quantum number $n$, as shown in figure \ref{potential}(a).
\begin{equation}  
	\Delta E_n = \frac{\pi^2}{2L^2}(2n+1).
	\label{eq:E_n}
\end{equation}

 To enable a quantitative comparison across different traps, we fix the lowest level spacing to be identical in all models ($\Delta E_1 = 1$). Consequently, the required well width is
\begin{equation}  
L_{\mathrm{ISW}} = \pi \sqrt{\frac{3}{2}}
	\approx 3.84.
	\label{eq:L_numeric}
\end{equation}

We also investigate the effect of decreasing level spacings on nonequilibrium dynamics, considering the one-dimensional hyperbolic Pöschl--Teller potential (PTP), defined in dimensionless form as
\begin{equation}
	V_{\mathrm{PTP}}(x) = - V_{0}\,\mathrm{sech}^{2}(a_{\mathrm{PTP}} x),
\end{equation}
where $a_{\mathrm{PTP}}$ controls the width of the well and $V_{0} > 0$ controls its depth. The corresponding Hamiltonian reads
\begin{equation}
	H_{\mathrm{PTP}}
	= -\frac{1}{2}\frac{\mathrm{d}^{2}}{\mathrm{d}x^{2}}
	- V_{0}\,\mathrm{sech}^{2}(a_{\mathrm{PTP}} x).
\end{equation}
The energies are given by
\begin{equation}
	E_{n} = -\frac{a_{\mathrm{PTP}}^{2}}{2}\,(\lambda - n)^{2},
	\qquad n = 0,1,\dots,n_{\max},
\end{equation}
where the highest bound state index satisfies $n_{\max} = \lfloor \lambda - 1 \rfloor$.
The corresponding normalized wavefunctions are
\begin{equation}
	\psi_{n}(x) 
	= \mathcal{N}_{n}\,
	\left[\mathrm{sech}(a_{\mathrm{PTP}} x)\right]^{\lambda - n}
	P_{n}^{(\lambda - n,\,\lambda - n)}\!\left[\tanh(a_{\mathrm{PTP}} x)\right],
\end{equation}
where $P_{n}^{(\alpha,\beta)}$ denotes the Jacobi polynomial and $\mathcal{N}_{n}$ is the normalization constant.

The first level spacing is
\begin{align}
	\Delta E_{0}
	&\equiv E_{1}-E_{0}=\frac{a_{\mathrm{PTP}}^{2}}{2}\,(2\lambda-1)=1,
\end{align}
therefore
\begin{equation}
	a_{\mathrm{PTP}} = \sqrt{\frac{2}{2\lambda-1}}.
\end{equation}
 The depth of the PTP is given by
\begin{equation}
V_0 = \frac{1}{2}\lambda(\lambda+1)a_{\mathrm{PTP}}^2.
\end{equation}
The single-particle spectrum takes the form $E_{n} = -\frac{1}{2\lambda-1}\,(\lambda - n)^{2}$. The corresponding level spacings are 
\begin{equation}
	\Delta E_{n}
	\equiv E_{n+1}-E_{n}
	= \frac{2(\lambda-n)-1}{2\lambda-1},
\end{equation}
which decrease monotonically and approximately linearly as $n$ increases, consistent with the compressed spectral structure shown in figure \ref{potential}(c). For the PTP, the spectral structure is controlled by the parameter $\lambda$, which determines both the number of bound states and the degree of spectral inhomogeneity\cite{Poschl-1933,Flugge-1994,daSilva-2024}. To investigate the role of spectral properties in nonequilibrium dynamics, we consider three paradigmatic one-dimensional trapping potentials: the infinite square well (ISW), the harmonic oscillator trap (HOT), and the Pöschl--Teller potential (PTP). As illustrated in figure \ref{potential}, these systems possess qualitatively distinct single-particle spectra. For clarity, only several low-lying energy levels are shown schematically. The ISW [figure \ref{potential}(a)] is characterized by increasing level spacings, the HOT [figure \ref{potential}(b)] exhibits an exactly equally spaced spectrum, and the PTP [figure \ref{potential}(c)] features progressively compressed level spacings. This classification provides a convenient framework for exploring how different spectral properties influence many-body dynamics.

For the PTP, the spectral properties are controlled by the parameter $\lambda$, which determines both the number of bound states and the degree of spectral inhomogeneity\cite{Poschl-1933,Flugge-1994,daSilva-2024}. As $\lambda$ increases, the variation of level spacings in the low-energy sector becomes progressively weaker, and the spectrum approaches the equally spaced limit. In the present work, we choose an intermediate value of $\lambda = 60$, which ensures a sufficiently large number of bound states while preserving the characteristic nonuniform spectral structure of the PTP\cite{Ahmed-2001}.

Each energy level is fourfold degenerate due to the spin components with $m = \pm1/2$ and $m = \pm3/2$. We consider an initially highly excited configuration in which only two particles occupy each energy level, specifically, only the $m=\pm1/2$ states are occupied. The many-body wave function $\Psi(\bm x)$ is constructed as a Slater determinant. Although four degenerate single-particle states exist at each level $n$, we consider only two of them $(s=\uparrow,\downarrow)$.

\begin{eqnarray}
	{\psi}_{n}^{s}(x) &=& \sum_{m=\pm3/2,\pm1/2} {\alpha}_{n,m}^{s}\varphi_n(x)\chi_m \notag\\
	&=&
	\sum_{m=\pm3/2,\pm1/2} \phi_{n,m}^{s}(x)\chi_m,
	\label{psi}
\end{eqnarray}
where ${\phi}_{n,m}^{s}(x)=\alpha_{n,m}^s\varphi_n(x)$ and $\alpha_{n,m}^s$ are the superposition coefficients for the four spin states. The function $\varphi_n(x)$ denotes the single-particle eigenfunction of the trapping potential (HOT, ISW, or PTP), and $\chi_m$ represents the spin basis.

The total energy of the system is
\begin{eqnarray}
	E&=&\int \Psi^{*}(\bm x)(H_0+H_I)\Psi(\bm x)d\bm x=E_0+E_I.
\end{eqnarray}
Here, $H_0$ is the single-particle Hamiltonian, and the interaction Hamiltonian is
\begin{eqnarray}
	H_I&=&\sum_{F=0}^{2f-1} \sum_{m_F=-F}^F g_F\delta(x-x')|F,m_F\rangle\langle F,m_F|.
\end{eqnarray}
 The reduced two-body interaction strength is defined as  $g_{F}={\tilde{g_F}}/({\hbar\omega \sqrt{\hbar/m\omega}})$, which is related to the $s$-wave scattering length $a_F$. The index $F$ labels the total spin channel ($F=0,1,2,3$). Since only $s$-wave scattering is considered, the spatial wavefunction of two identical fermions is symmetric, and the spin wavefunction must be antisymmetric. Consequently, the interaction is completely characterized by two coupling constants, $g_0$ and $g_2$, corresponding to the even scattering channels $F=0$ and $F=2$, respectively.

The TDHF equations are obtained by varying the action $(E - \mathrm{i}\hbar\partial_t N)$ to $\psi_{n}^{s}$, 

\begin{eqnarray}
	(\mathrm{i}\hbar\partial_t - H_0)\psi_{n}^{s}(x) = \sum_{n^{\prime}, s^{\prime}} {\bm M}_{n^{\prime}}^{s^{\prime}}\psi_{n}^{s}(x).
\end{eqnarray}
 Within the TDHF approximation, interaction effects enter through self-consistent mean fields, which preserve the coherent spin-exchange processes responsible for spin-mixing dynamics. This approach has been widely used to describe nonequilibrium dynamics in weakly interacting multicomponent Fermi gases\cite{FengT-2017,Lisy-2025,Lisy-2026}. In the present work, we consider finite trapped systems with discrete bound states and focus on the weak-interaction regime, where the interaction energy remains smaller than the characteristic single-particle level spacing. Under these conditions, the dynamics is expected to be dominated by coherent spin-exchange processes among a finite number of trapped orbitals, for which a mean-field description provides a reasonable approximation.
 
The matrix ${\bm M}_{n^{\prime}}^{s^{\prime}}$ describes the self-consistent mean-field interaction effects. In particular, its off-diagonal elements generate coherent couplings among different Zeeman components and thereby retain the coherent spin-exchange dynamics within the TDHF framework. Although correlations beyond the mean-field level are neglected, the self-consistent mean fields provide a consistent description of the coherent spin-mixing dynamics considered in the present work. The parameter $c_2$, originating from the difference between the two scattering channels ($g_2-g_0$), characterizes the spin-exchange interaction and coherently couples the $m=\pm1/2$ and $m=\pm3/2$ components, thereby enabling spin-mixing dynamics. In the absence of this term ($c_2=0$), spin-exchange processes vanish and no spin mixing occurs. The parameter $c_0$, originating from $g_2+g_0$, describes the non-spin-exchange interaction. Throughout this work, $c_2$ is fixed at $0.1$ to ensure finite spin mixing, whereas $c_0$ is varied to examine how the decay behavior depends on the interaction strength. For convenience, all variables are set to be dimensionless.

\begin{widetext}
	\begin{eqnarray}
		\centering
		{\bm M}_{n^\prime}^{s^\prime}&=&\renewcommand{\arraystretch}{1.2}\setlength{\arraycolsep}{4pt}
	\left(\begin{matrix}
		c_0|\phi_{n^{\prime},-3/2}^{s^{\prime}}|^{2} &
		(c_2-c_0){{M_1}}&
		-(c_2+c_0){M_2}&
		-c_0{M_3} \\
		(c_2-c_0){{M_1}^\ast} &
		c_0|\phi_{n^{\prime},-1/2}^{s^{\prime}}|^{2} &
		-c_0{M_4}  &
		-(c_2+c_0){M_5} \\
		-(c_2+c_0){{M_2}^\ast}  &
		-c_0{{M_4}^\ast}  &
		c_0|\phi_{n^{\prime},1/2}^{s^{\prime}}|^{2} &
		(c_2-c_0){M_6} \\
		-c_0{{M_3}^\ast} &
		-(c_2+c_0){{M_5}^\ast} &
		(c_2-c_0){{M_6}^\ast} &
		c_0|\phi_{n^{\prime},3/2}^{s^{\prime}}|^{2}
	\end{matrix}\right),
	\label{M}
	\end{eqnarray}
\end{widetext}

with
\begin{align}
	M_1 &= \phi_{n',-3/2}^{s',*} \phi_{n',-1/2}^{s'}, &
	M_2 &= \phi_{n',-3/2}^{s',*} \phi_{n', 1/2}^{s'}, \nonumber \\
	M_3 &= \phi_{n',-3/2}^{s',*} \phi_{n', 3/2}^{s'}, &
	M_4 &= \phi_{n',-1/2}^{s',*} \phi_{n', 1/2}^{s'},  \nonumber \\
	M_5 &= \phi_{n',-1/2}^{s',*} \phi_{n', 3/2}^{s'}, &
	M_6 &= \phi_{n', 1/2}^{s',*} \phi_{n', 3/2}^{s'}.
\end{align}

\section{Results and discussion}

We prepare an initial state where each energy level is occupied by two particles. The corresponding single-particle wave functions are chosen as
\begin{eqnarray}
	\psi_{n}^{\uparrow}(x) &=& \mathrm{e}^{-\mathrm{i}S_x\theta}[0,1,0,0]^{T}\varphi_n(x), \notag\\
	\psi_{n}^{\downarrow}(x) &=& \mathrm{e}^{-\mathrm{i}S_x\theta}[0,0,1,0]^{T}\varphi_n(x).
\end{eqnarray}
$S_x$ is the spin matrix along the $x$ direction, and $\theta=0.05$ is a small rotation angle that controls the initial spin superposition. This operation introduces weak mixing among the spin-$z$ eigenstates while keeping the population predominantly concentrated in the $m = \pm1/2$ manifold. Under this preparation, the relative phases between spin components and orbitals are fixed. The many-body initial state $\Psi(0)$ is then constructed as a Slater determinant of these single-particle wave functions. With time evolution, the spin-mixing interaction ($c_2 \neq0$) coherently couples the $m = \pm1/2$ and $m = \pm3/2$ components, leading to oscillatory population transfer between them.

\begin{figure}
	\begin{center}
		\includegraphics[width=0.95\linewidth]{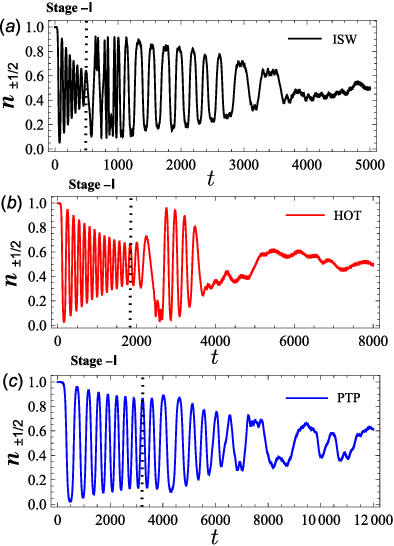}
	\end{center}
	\caption{\label{long time oscillation} {\bf The long-time relative populations as a function of time for different traps.} The relative population $n_{\pm1/2}$ (for the $m=\pm 1/2$ components) is shown as a function of time. (a) corresponds to the infinite square well, (b) to the harmonic oscillator trap, and (c) to the Pöschl-Teller potential. Stage-I is the regime corresponding to the regular oscillation. The total particle number is $N=20$, with interaction strengths $c_2=0.1$ and $c_0=1.0$.}
\end{figure}

\begin{figure*}
	\begin{center}
		\includegraphics[width=0.9\linewidth]{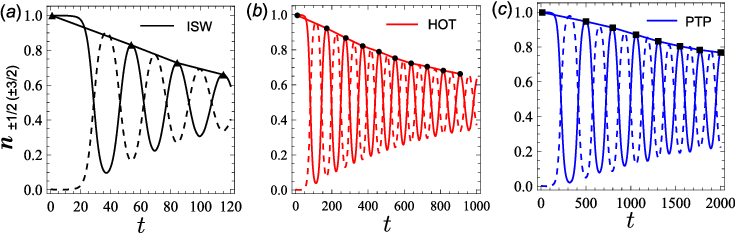}
	\end{center}
	\caption{\label{oscillation} {\bf Spin-mixing dynamics and envelope decay for different traps.} $n_{\pm 1/2}$ (solid lines) and $n_{\pm 3/2}$ (dashed lines) represent the relative populations of the $m = \pm 1/2$ and $m = \pm 3/2$ spin components, respectively. The time evolution of the peak values of $n_{\pm 1/2}(t)$, denoted as $A(t)$ and shown by symbols, is well fitted with a power-law function, $A(t)=A_0-\gamma t^{\alpha}$. The total particle number is $N=40$, with interaction parameters $c_2=0.1$ and $c_0=1.0$.}
\end{figure*}

\begin{figure}
	\begin{center}
		\includegraphics[width=0.95\linewidth]{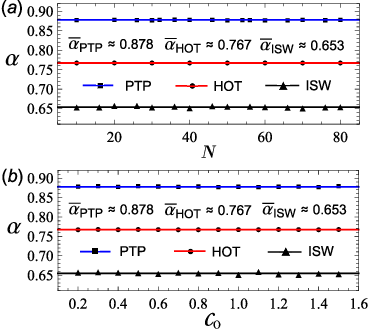}
	\end{center}
	\caption{\label{alphaNC} {\bf The exponent $\alpha$ extracted from the envelope of the dynamics as a function of particle number $N$ and interaction strength  $c_{0}$.} Here, $N$ denotes the total particle number, and $c_{0}$ characterizes the interaction strength. For each trapping potential, the exponent $\alpha$ remains essentially independent of both $N$ and $c_{0}$.}
\end{figure}

\begin{figure*}
	\begin{center}
		\includegraphics[width=0.9\linewidth]{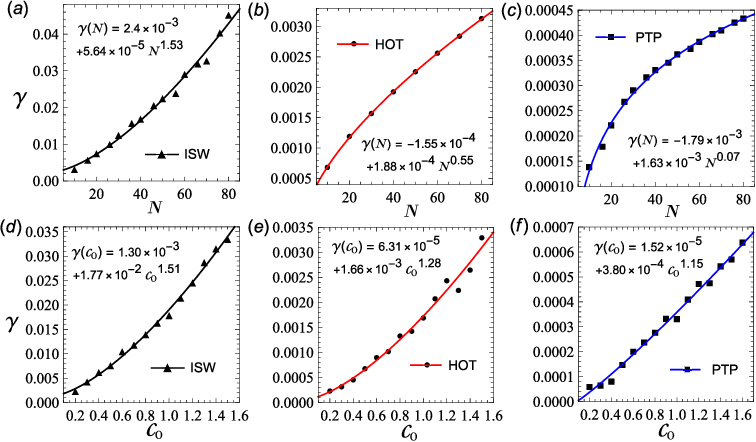}
	\end{center}
	\caption{\label{gammaNC} {\bf Decay parameter $\gamma$ as a function of particle number $N$ and interaction strength  $c_{0}$.} (a)-(c) $\gamma$ increases monotonically with $N$, but the functional dependence of $\gamma$ on $N$ differs among the three traps. The interaction strengths are $c_2=0.1$ and $c_0=1.0$. (d)-(f) $\gamma$ increases monotonically with $c_{0}$, indicating that stronger interactions enhance the damping of the oscillations. The total particle number is $N=40$.}
\end{figure*}

For clarity, figure \ref{long time oscillation} shows only the long-time evolution of the relative population $n_{\pm 1/2}$. In our previous work\cite{Lisy-2025}, we showed that, in a harmonic oscillator trap (HOT), the spin-mixing dynamics exhibits a two-stage behavior: Stage-I is characterized by regular oscillations with slowly decaying envelopes, followed by a more complex regime accompanied by partial revivals. As shown in figure \ref{long time oscillation}, all three traps exhibit persistent oscillatory behavior over long times and eventually approach quasi-stationary states. Although a two-stage dynamical structure is observed in all cases, its manifestation depends on the trapping potential. The ISW exhibits revival patterns similar to those found in the HOT, whereas in the PTP the oscillations become increasingly irregular at long times and large-amplitude revivals are not evident. In addition, the characteristic timescale for approaching the quasi-stationary regime differs significantly among the three traps: the ISW evolves most rapidly, the PTP exhibits the slowest evolution, and the HOT lies in between. Despite the different long-time behaviors, all three systems exhibit regular oscillations with slowly decaying envelopes during Stage I. We therefore focus on this regime and extract the decay behavior from the oscillation envelopes. Within the present TDHF framework, the reduction of the oscillation amplitude during Stage I can be qualitatively understood primarily in terms of coherent dephasing among dynamically participating modes rather than irreversible dissipation. As shown in figure \ref{oscillation}, the ISW displays the fastest decay of oscillations, whereas the PTP supports the longest-lived coherent dynamics.

The envelope of the oscillations can be well described by a power-law function
\begin{equation}
	A(t) = A_{0} - \gamma t^{\alpha},
\end{equation}
which provides an excellent description of the decay behavior across all three trapping potentials considered here. For convenience, $A(t)$ is referred to as the oscillation amplitude, and $A_0$ denotes its initial value. This parametrization enables us to characterize the dynamics in terms of the exponent $\alpha$ and the decay parameter $\gamma$, whose dependence on system parameters will be discussed below.

The decay exponent $\alpha$ extracted from the oscillation envelopes is shown in figure \ref{alphaNC} as a function of particle number $N$ (a) and interaction strength $c_{0}$ (b). For each trapping potential, $\alpha$ remains nearly constant over wide ranges of both $N$ and $c_{0}$, indicating only weak sensitivity to these parameters. In contrast, $\alpha$ differs systematically among the three trapping potentials. Specifically, $\alpha$ is largest for the PTP ($\bar{\alpha}_{\mathrm{PTP}}\approx0.878$), intermediate for the HOT ($\bar{\alpha}_{\mathrm{HOT}}\approx0.767$), and smallest for the ISW ($\bar{\alpha}_{\mathrm{ISW}}\approx0.653$). Since these trapping potentials possess qualitatively different single-particle spectral structures, the systematic variation of $\alpha$ suggests a close connection between the decay dynamics and the underlying spectrum. The robustness of $\alpha$
against variations in $N$ and $c_{0}$, together with its systematic
dependence on the spectral structure, supports the view that the power-law
form captures a robust feature of the decay dynamics rather than merely
describing an individual data set.

A possible qualitative interpretation is based on coherent dephasing. The spin dynamics involves the coherent superposition of multiple dynamical modes associated with different single-particle orbitals and interaction-induced couplings. These modes generally accumulate different phases during the time evolution, leading to progressive phase dispersion and a reduction of
the coherent oscillation amplitude. Since the single-particle
spectrum influences the distribution of characteristic frequencies,
different trapping potentials can therefore exhibit different dephasing
dynamics, providing a possible explanation for the observed variation of the
decay exponent $\alpha$. In particular, the increasingly separated level spacings of the ISW can enhance the dispersion of characteristic frequencies and thereby accelerate dephasing, whereas the progressively compressed spectrum of the PTP tends to suppress such phase dispersion. This picture is consistent with the smaller
$\alpha$ observed for the ISW and the larger $\alpha$ observed for the PTP. We emphasize that this interpretation provides a qualitative physical picture rather than an analytical derivation of the power-law form. A quantitative relation
between the decay exponent and specific properties of the single-particle
spectrum remains to be established.

Figure \ref{gammaNC} (a)-(c) shows the dependence of $\gamma$ on particle number $N$ for the ISW, HOT, and PTP. In all three cases, $\gamma$ increases monotonically with $N$, indicating that larger systems exhibit faster decay of coherent oscillations. At the same time, the magnitude of $\gamma$ differs significantly among the three traps. For comparable particle numbers, the ISW consistently exhibits the largest $\gamma$, the PTP the smallest, with the HOT lying in between. A similar trend is observed in the dependence of $\gamma$ on the interaction strength $c_{0}$, as shown in figure \ref{gammaNC} (d)-(f). For all three traps, increasing $c_{0}$ leads to a monotonic increase in $\gamma$,  indicating that stronger interactions accelerate the decay of oscillations. Importantly, the relative ordering of the decay parameters remains unchanged as $c_{0}$ varies: the ISW always exhibits the largest $\gamma$, followed by the HOT, while the PTP remains the smallest.

In the preceding analysis, the first level spacing was fixed across different traps, enabling a controlled comparison of distinct spectral structures. However, this constraint also fixes the overall energy scale. To disentangle these contributions, we vary the trap width, thereby rescaling the overall energy scale while preserving the relative structure of the spectrum. This allows us to separately examine the roles played by the spectral structure and the overall energy scale in the dynamics.

\begin{figure}[!t]
	\begin{center}
		\includegraphics[width=0.95\linewidth]{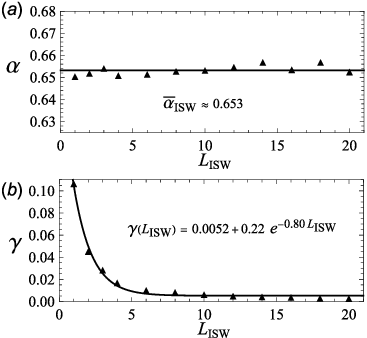}
	\end{center}
	\caption{\label{alphaLISW} {\bf Dependence of the decay exponent $\alpha$ and decay parameter $\gamma$ on the well width $L_{\mathrm{ISW}}$ for the infinite square well.} (a) shows that the exponent $\alpha$ remains nearly constant as the well width varies, whereas (b) shows a strong suppression of the decay parameter $\gamma$ with increasing $L_{\mathrm{ISW}}$. The particle number is $N=40$, and the interaction strengths are $c_2=0.1$ and $c_0=1.0$.}
\end{figure}

\begin{figure}
	\begin{center}
		\includegraphics[width=0.95\linewidth]{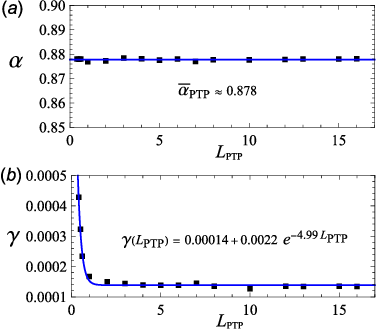}
	\end{center}
	\caption{\label{alphaLPT} {\bf Dependence of the decay exponent $\alpha$ and decay parameter $\gamma$ on the effective width $L_{\mathrm{PTP}}$.} The exponent $\alpha$ remains essentially unchanged as $L_{\mathrm{PTP}}$ varies, whereas the decay parameter $\gamma$ decreases rapidly as the trap becomes wider, indicating enhanced stability of coherent oscillations in broader PTP. The particle number is $N=40$, and the interaction strengths are $c_2=0.1$ and $c_0=1.0$.}
\end{figure}

In figure \ref{alphaLISW}, we relax the constraint of fixed level spacing and directly vary the well width $L_{\mathrm{ISW}}$ in the infinite square well. As shown in figure \ref{alphaLISW}(a), the decay exponent $\alpha$ remains nearly unchanged over a wide range of $L_{\mathrm{ISW}}$, indicating that it is largely insensitive to the overall energy scale of the spectrum. In contrast, figure \ref{alphaLISW}(b) shows a pronounced dependence of the decay parameter $\gamma$ on $L_{\mathrm{ISW}}$. As the well width increases, $\gamma$ decreases rapidly, indicating a significant slowing down of the oscillation decay. This behavior can be understood from the scaling of the single-particle spectrum in the infinite square well, where energy levels scale as $E_n \propto n^2 / L_{\mathrm{ISW}}^2$. Increasing $L_{\mathrm{ISW}}$ reduces the overall level spacing, thereby decreasing the spread of dynamical phases accumulated by different modes and leading to slower dephasing.

Before analyzing the dependence of the nonequilibrium dynamics on the effective width of the PTP, it is necessary to clarify the choice of potential parameters. The one-dimensional PTP is written as
\begin{equation}
V_{\mathrm{PTP}}(x) = -\frac{1}{2}\,\lambda(\lambda+1)\,a_{\mathrm{PTP}}^{2}\,
\mathrm{sech}^{2}\!\left(a_{\mathrm{PTP}} x\right),
\end{equation}
where $a_{\mathrm{PTP}}$ sets the inverse length scale of the confinement. To characterize the spatial extent of the trap, we introduce its effective width
\begin{equation}
	L_{\mathrm{PTP}} \equiv \frac{1}{a_{\mathrm{PTP}}}.
\end{equation}
With this definition, decreasing $a_{\mathrm{PTP}}$ corresponds to a wider trap, while increasing $a_{\mathrm{PTP}}$ leads to a tighter confinement. For the PTP, the energy spectrum is $E_n = -\frac{a_{\mathrm{PTP}}^{2}}{2}\,(\lambda-n)^2, n=0,1,\dots,\lfloor\lambda-1\rfloor.$ Therefore, the spacing between neighboring energy levels is 
\begin{equation}
	\Delta E_n \equiv E_{n+1}-E_n
	= \frac{2(\lambda-n)-1}{2\,L_{\mathrm{PTP}}^{2}}.
\end{equation}
This expression explicitly shows that increasing $L_{\mathrm{PTP}}$ leads to a global reduction in energy-level separations and a compression of the energy spectrum. The effective width $L_{\mathrm{PTP}}$ is varied while keeping $\lambda$ fixed at $60$. Under these conditions, changing $L_{\mathrm{PTP}}$ mainly induces a global rescaling of the energy spectrum while preserving its qualitative structure within the relevant low-energy sector.

Figure \ref{alphaLPT} shows the corresponding dependence of the decay behavior on the effective trap width for the PTP. Similar to the ISW case, the decay exponent $\alpha$ remains nearly constant over a wide range of $L_{\mathrm{PTP}}$, further confirming its insensitivity to the overall energy scale. By contrast, the decay parameter $\gamma$ exhibits a pronounced dependence on $L_{\mathrm{PTP}}$, decreasing rapidly as the potential becomes wider. This trend can be qualitatively understood in terms of dephasing. Increasing $L_{\mathrm{PTP}}$ reduces the energy-level spacings and narrows the distribution of dynamical frequencies, thereby slowing down the dephasing process. As a result, coherent oscillations persist over longer timescales in wider PTPs. The consistent behavior observed in figures \ref{alphaLISW} and \ref{alphaLPT} leads to a general conclusion: geometric parameter variations that primarily rescale the energy spectrum strongly affect the decay parameter $\gamma$, while leaving the decay exponent $\alpha$ largely unchanged.

\section{Conclusion}

In this work, we have investigated how distinct single-particle spectral structures influence the nonequilibrium spin dynamics of trapped spin-$3/2$ Fermi gases beyond the previously studied harmonic trap. The dynamics exhibits a two-stage structure: an initial regime characterized by regular oscillations with a slowly decaying envelope (Stage I), followed by a long-time regime displaying increasingly complex behavior. Since the long-time dynamics becomes irregular, we focus on Stage I, where the decay process can be characterized in a controlled and quantitative manner. Within this regime, we find that the decay of coherent oscillations follows a power-law form despite pronounced differences in their underlying single-particle spectra. The exponent $\alpha$ is largely insensitive to particle number and interaction strength, but exhibits systematic variations among the three trapping potentials, taking the largest value in the PTP, an intermediate value in the HOT, and the smallest value in the ISW. In contrast, the decay parameter $\gamma$ depends strongly on the spectral structure. By varying the effective trap width, we further show that changes in the overall energy scale primarily affect $\gamma$, while leaving $\alpha$ essentially unchanged. These results show that power-law decay is not restricted to the harmonic trap but persists across qualitatively distinct spectral structures. The systematic variation of the decay exponent among different traps suggests a close connection between the decay behavior and the underlying single-particle spectrum. A possible qualitative interpretation is that different spectral structures are expected to produce distinct distributions of dynamical frequencies and consequently different dephasing behaviors, which may contribute to the observed variations in the decay characteristics. More generally, our findings highlight the importance of spectral structure in nonequilibrium dynamics and suggest that confinement geometry provides an effective route for controlling decay processes in interacting high-spin Fermi gases.

\textbf{Acknowledgments}\\
 This work was supported by the National Natural Science Foundation of China (Grant Nos. 11574028 and 11874083).
\\
\textbf{Data availability}\\
Data are available upon request.
\\
\textbf{Author contributions}\\
$\bf{Shuyi Li:}$ Investigation, Conceptualization, Methodology, Data curation, Writing - original draft. $\bf{Qiang Gu:}$  (Corresponding Author) Conceptualization, Supervision, Writing - reviewing.
\\
\textbf{Declaration of Interest Statement}\\
The authors declared that they have no conflicts of interest related to this work. They do not have any commercial or associative interest that represents a conflict of interest in connection with the work submitted. 
\\


\begin{thebibliography}{99}
	
	\bibitem{K.G.Wilson-1975} K. G. Wilson, The renormalization group: Critical phenomena and the Kondo problem, Rev. Mod. Phys. \textbf{47} (1975) 773.
	
	\bibitem{K.G.Wilson-1983} K. G. Wilson, The renormalization group and critical phenomena, Rev. Mod. Phys. \textbf{55} (1983) 583.
	
	\bibitem{Erne-2018} S. Erne, R. Bücker, T. Gasenzer, J. Berges, and  J. Schmiedmayer, Universal dynamics in an isolated one-dimensional Bose gas far from equilibrium, Nature \textbf{563} (2018) 225.
	
	\bibitem{Prufer-2018}  M. Prüfer, P. Kunkel, H. Strobel, S. Lannig, D. Linnemann, C. M. Schmied, J. Berges, T. Gasenzer, and M. K. Oberthaler, Observation of universal dynamics in a spinor Bose gas far from equilibrium, Nature \textbf{563} (2018) 217.
	
	\bibitem{Pecci-2022} G. Pecci, P. Vignolo, and A. Minguzzi, Universal spin-mixing oscillations in a strongly interacting one-dimensional Fermi gas, Phys. Rev. A \textbf{105} (2022) 051303.
	
	\bibitem{Rodriguez-2022} J. F. Rodriguez-Nieva, A. P. Orioli, and J. Marino, Far-from-equilibrium universality in the two-dimensional Heisenberg model, Proc. Natl. Acad. Sci. \textbf{119} (2022) e2122599119.
	
	\bibitem{ZhaoyiLi-2023} Z. Li, P. Glorioso, and J. F. Rodriguez-Nieva, Far-from-equilibrium universality in two-dimensional Heisenberg antiferromagnets, Phys. Rev. B \textbf{108} (2023) 144303.
	
	\bibitem{LiY-2024} Y. Li, T. Zhou, Z. Wu, P. Peng, S. Zhang, R. Fu, R. Zhang, W. Zheng, P. Zhang, H. Zhai, X. Peng, and J. Du, Emergent universal quench dynamics in randomly interacting spin models, Nat. Phys. \textbf{20} (2024) 1966.
	
	\bibitem{Bloch-2008} I. Bloch, J. Dalibard, and W. Zwerger, Many-body physics with ultracold gases, Rev. Mod. Phys. \textbf{80} (2008) 885.
	
	\bibitem{Chin-2010} C. Chin, R. Grimm, P. Julienne, and E. Tiesinga, Feshbach resonances in ultracold gases, Rev. Mod. Phys. \textbf{82} (2010) 1225.
	
	\bibitem{Polkovnikov-2011} A. Polkovnikov, K. Sengupta, A. Silva, and M. Vengalattore, Colloquium: Nonequilibrium dynamics of closed interacting quantum systems, Rev. Mod. Phys. \textbf{83} (2011) 863.
	
	\bibitem{Ueda-2020} M. Ueda, Quantum equilibration, thermalization and prethermalization in ultracold atoms, Nat. Rev. Phys. \textbf{2} (2020) 669.
	
	\bibitem{Stringari-1996} S. Stringari, Collective excitations of a trapped Bose-condensed gas, Phys. Rev. Lett. \textbf{77} (1996) 2360.
	
	\bibitem{Vignolo-2000} P. Vignolo, A. Minguzzi, and M. P. Tosi, Exact particle and kinetic-energy densities for one-dimensional confined gases of noninteracting Fermions, Phys. Rev. Lett. \textbf{85} (2000) 2850.
	
	\bibitem{Minguzzi-2002} A. Minguzzi, P. Vignolo, and M. P. Tosi, High-momentum tail in the Tonks gas under harmonic confinement, Phys. Lett. A \textbf{294} (2002) 222.
	
    \bibitem{Berry-1977} M. V. Berry, Regular and irregular semiclassical wavefunctions, J. Phys. A \textbf{10} (1977) 2083.
    
    \bibitem{Robinett-2004} R. W. Robinett, Quantum wave packet revivals, Phys. Rep. \textbf{392} (2004) 1.
	
	\bibitem{Krauser-2014} J. S. Krauser, U. Ebling, N. Fläschner, J. Heinze, K. Sengstock, M. Lewenstein, A. Eckardt, and C. Becker, Giant spin oscillations in an ultracold Fermi sea, Science \textbf{343} (2014) 157.
	
	\bibitem{Dong-2013} Y. Dong and H. Pu, Spin mixing in spinor Fermi gases, Phys. Rev. A \textbf{87} (2013) 043610.
	
	\bibitem{FengT-2017} T. Feng, J. Xu, and Q. Gu, Spin dynamics of large-spin fermions in a harmonic trap, Ann. Phys. \textbf{377} (2017) 175.
	
	\bibitem{Lisy-2025} S. Li and Q. Gu, Universal behaviors of dynamical quantum transition in trapped large-spin fermions, Ann. Phys. \textbf{475} (2025) 169952.
	
	\bibitem{Lisy-2026} S. Li and Q. Gu, Nonthermal dynamics and scarlike spectral structures in a high-spin Fermi gas, Phys. Rev. B \textbf{113} (2026) 104303.
	
	\bibitem{Krauser-2012} J. S. Krauser, J. Heinze, N. Fläschner, et al., Coherent multi-flavour spin dynamics in a fermionic quantum gas, Nat. Phys. \textbf{8} (2012) 813.
	
    \bibitem{Heinze-2013} J. Heinze, J. S. Krauser, N. Fläschner,
    K. Sengstock, C. Becker, U. Ebling,
    A. Eckardt, and M. Lewenstein, Engineering spin waves in a high-spin ultracold Fermi gas, Phys. Rev. Lett. \textbf{110} (2013) 250402.
	
	\bibitem{Ebling-2014} U. Ebling, J. S. Krauser, N. Fläschner, K. Sengstock, C. Becker, M. Lewenstein, and A. Eckardt, Relaxation dynamics of an isolated large-spin Fermi gas far from equilibrium, Phys. Rev. X \textbf{4} (2014) 021011.
	
	\bibitem{Poschl-1933} G. Pöschl and E. Teller, Bemerkungen zur Quantenmechanik des anharmonischen Oszillators, Z. Phys. \textbf{83} (1933) 143.
	
	\bibitem{Flugge-1994} S. Fl\"ugge, Practical Quantum Mechanics, Springer, Berlin (1994), p. 88.
	
	\bibitem{Lekner-2007} J. Lekner, Reflectionless eigenstates of the sech$^2$ potential, Am. J. Phys. \textbf{75} (2007) 1151.
	
	\bibitem{daSilva-2024} U. C. da Silva, C. F. S. Pereira, and A. A. Lima, Renormalization group and spectra of the generalized Pöschl–Teller potential, Ann. Phys. \textbf{460} (2024) 169549.
	
	\bibitem{Ahmed-2001} Z. Ahmed, Real and complex discrete eigenvalues in an exactly solvable one-dimensional complex PT-invariant potential, Phys. Lett. A \textbf{282} (2001) 343.
	
\end{thebibliography}
\end{document}